\documentclass[letterpaper, 10pt, conference]{ieeeconf}
\IEEEoverridecommandlockouts
\usepackage[T1]{fontenc}

\usepackage{comment}

\usepackage[sort,compress]{cite}

\usepackage{amssymb}
\usepackage{mathtools}
\usepackage{etoolbox}
\usepackage{bm}
\let\labelindent\relax
\usepackage[inline]{enumitem}

\usepackage[dvipsnames, cmyk]{xcolor}
\definecolor{myred}{HTML}{d7145b}
\definecolor{myblue}{HTML}{1e88e5}
\definecolor{myyellow}{HTML}{ff8500}
\definecolor{mygreen}{HTML}{34fc56}
\definecolor{myaqua}{HTML}{24dc96}
\definecolor{mypurple}{HTML}{EF3720}

\usepackage{graphicx, tikz, pgfmath, pgfplots}
\usepackage{subcaption, standalone}
\usepgfplotslibrary{fillbetween}
\usepgfplotslibrary{statistics}
\pgfplotsset{compat=1.18}
\usetikzlibrary{positioning,calc}
\usetikzlibrary{arrows.meta}
\tikzset{>={Latex[width=2mm,length=1.5mm]}}
\pgfmathsetseed{10}%

\pgfplotsset{%
	compat=1.18,
    set layers,
	my boxplot/.style={
		boxplot,
		mark=x,
		boxplot/every box/.style={
			solid,
			draw=Blue,
			fill=White,
			line width=0.3mm,
		},
		boxplot/every whisker/.style={
			solid,
			draw=Black,
			line width=0.3mm,
		},
		boxplot/every median/.style={
			solid,
			draw=BrickRed,
			line width=0.3mm,
		},
		every mark/.style={
			draw=Red,
			mark size=2pt,
			line width=0.1mm,
		},
	}
}

\newcommand{\st}{\,:\,}

\DeclarePairedDelimiterX\RVvar[1]{.}{.}{
	 #1
}

\DeclarePairedDelimiterXPP\RVpdf[2]{#1}{\lparen}{\rparen}{}{
	 #2
}

\DeclarePairedDelimiterXPP\RVexpected[1]{\mathbb{E}}{\lbrack}{\rbrack}{}{
	 #1
}

\DeclarePairedDelimiterXPP\RVprob[1]{\mathbb{P}}{\lbrace}{\rbrace}{}{
	 #1
}

\DeclarePairedDelimiterX\norm[1]{\lVert}{\rVert}{
	\ifblank{#1}{\,\cdot\,}{#1}
}

\DeclarePairedDelimiterX\inner[2]{\langle}{\rangle}{
	\ifblank{#1}{\,\cdot\,}{#1},
	\ifblank{#2}{\,\cdot\,}{#2}
}

\DeclarePairedDelimiterX\defVector[2]{[}{]}{ #1 \st #2 }
\DeclarePairedDelimiterX\defMatrix[3]{[}{]}{ #1 \!\! \st \!\! \begin{cases*} #2 \\ #3 \end{cases*} }

\title{\LARGE \bf
eXplainable AI for data driven control:\\ an inverse optimal control approach
}

\author{%
	Federico~Porcari$^{1}$, Donatello~Materassi$^{2}$, Simone~Formentin$^{3}$
	\thanks{This work is partially supported by the FAIR project (NextGenerationEU, PNRR-PE-AI, M4C2, Investment 1.3), the 4DDS project (Italian Ministry of Enterprises and Made in Italy, grant F/310097/01-04/X56), and the PRIN PNRR project P2022NB77E (NextGenerationEU, CUP: D53D23016100001). It is also partly supported by the ENFIELD project (Horizon Europe, grant 101120657).	}
	\thanks{$^1$%
		Federico Porcari is with the Department of Electronics, Information and Bioengineering, Politecnico di Milano, via G. Ponzio 34/5, 20133 Milano, Italy.
		Email: \texttt{federico.porcari@polimi.it}
	}%
    	\thanks{$^3$%
		Donatello Materassi is with Department of Electrical and Computer Engineering, University of Minnesota, Twin Cities.
		Email: \texttt{mater013@umn.edu}
	}%
	\thanks{$^3$%
		Simone Formentin is with the Department of Electronics, Information and Bioengineering, Politecnico di Milano, via G. Ponzio 34/5, 20133 Milano, Italy.
		Email: \texttt{simone.formentin@polimi.it}
	}%
}

\begin{document}

\maketitle

\begin{abstract}
    Understanding the behavior of black-box data-driven controllers is a key challenge in modern control design. In this work, we propose an eXplainable AI (XAI) methodology based on Inverse Optimal Control (IOC) to obtain local explanations for the behavior of a controller operating around a given region. 
    Specifically, we extract the weights assigned to tracking errors and control effort in the implicit cost function that a black-box controller is optimizing, offering a more transparent and interpretable representation of the controller’s underlying objectives.
    This approach presents connections with well-established XAI techniques, such as Local Interpretable Model-agnostic Explanations (LIME) since it is still based on a local approximation of the control policy. 
    However, rather being limited to a standard sensitivity analysis, the explanation provided by our method relies on the solution of an inverse Linear Quadratic (LQ) problem, offering a structured and more control-relevant perspective.
    Numerical examples demonstrate that the inferred cost function consistently provides a deeper understanding of the controller’s decision-making process, shedding light on otherwise counterintuitive or unexpected phenomena.
\end{abstract}
\smallskip

\section{Introduction}\label{sec:intro}
AI has been advancing rapidly, leading to its application across a wide array of domains, from healthcare and finance to manufacturing and transportation \cite{chen2021survey,cao2022ai,li2023artificial,bi2023comprehensive,hamet2017artificial}.
However, a major problem with AI-based algorithms is the lack of transparency in their decision-making processes \cite{ryan2020ai,ribeiro2016should}, which has spurred the development of a new research field called eXplainable Artificial Intelligence (XAI), specifically focused on creating methods to interpret how these systems make decisions \cite{schwalbe2023comprehensive,dwivedi2023explainable,rong2023towards}. 
Despite the growing interest for XAI methods, there is no clear consensus or formalization on what constitutes an ``explanation.'' 
Currently, the most common distinction in XAI methods is between global explanations, which seek to provide a broad understanding of the model’s behavior across the entire input space, and local explanations, which focus on analyzing the model’s decisions for a specific instance or trajectory, offering insights into why a particular outcome was produced in a given scenario \cite{minh2022explainable}.
Common global explanation techniques include approximations using decision trees \cite{sivaprasad2023evaluation}, rule-based models \cite{van2021evaluating}, and Accumulated Local Effects (ALE) along with its variations \cite{apley2020visualizing,gkolemis2023dale}; while prevalent local explanation methods include LIME (Local Interpretable Model-Agnostic Explanations) \cite{ribeiro2016should}, SHAP (SHapley Additive exPlanations) \cite{lundberg2017unified,biparva2024incorporating},  counterfactual explanations \cite{mothilal2020explaining,guidotti2022counterfactual}, and saliency maps.
Typically, local explanations are easier to obtain, as they require analyzing only a limited region of the input space rather than the entire model. Despite their localized scope, they remain highly valuable, particularly for debugging purposes and individual failure analysis. Additionally, local explanations are often easier to categorize, as they frequently belong to the class of feature attribution methods \cite{lundberg2017unified}. 
In these methods, the explanation takes the form of a vector, with each entry indicating the contribution of an individual input component to the output, offering a structured and direct way to assess input relevance.

More recently, AI applications have also been extending into the design of control algorithms, where AI techniques are leveraged to improve the performance of self-operating machines such as drones, self-driving cars, and robotic systems \cite{zhang2020autonomous,ma2020artificial}.
The lack of transparency in AI-based control algorithms is especially problematic for autonomous systems, since it presents very specific challenges not satisfactorily addressed by current state-of-the-art XAI methods.
Indeed, control systems typically operate through a dynamic process where decisions are not made in isolation; instead, they take decisions in real-time often solving constrained optimization problems over an extended time horizon, and operating within a feedback loop alongside the system under control.
All these characteristics add increased complexity to understanding their behavior and driving increased research interest in developing dedicated methods for interpretable control \cite{biparva2023application,biparva2023interpretation,riva2024towards,allamaa2025exampcdatadrivenexplainableapproximate}.
A significant challenge in applying XAI methods to control problems is precisely that most existing XAI techniques are designed to explain static quantities, such as individual predictions in one-shot classification or regression tasks.
Unlike static features, which have well-defined individual values, control systems operate over time, generating time series data that embody complex temporal dependencies and evolving patterns \cite{materassi2024explaining,materassi2024explaining2}.

The main goal of this article is to develop a local explainability tool, that is specifically designed for control systems. 
Unlike standard XAI methods, which typically focus on estimating how much a single static feature influences the output, our approach aims to assess how individual temporal signals affect the system’s behavior over time.
This is achieved by leveraging specific control-theoretical tools.
Given a black-box controller and a particular trajectory to explain, we obtain a local explanation through a linearization procedure around the trajectory to be explained, followed by solving an inverse optimal control (IOC) problem.
The cost function derived from the inverse optimal control process serves as the explanation, as its knowledge contains information about the relative importance of each input and output time series in terms of control effort and performance.
This provides insights into what the black-box controller is effectively optimizing, shedding light on the implicit trade-offs guiding its decision-making process.

Interestingly, this specific explanation tool can be encapsulated within a more general framework that enables a more formal definition of explanations. Under this framework, the core principle is to view an explanation as \textit{a form of reduced complexity model} \cite{materassi2009reduced}. 
Like many existing XAI methods, our approach recognizes that explanations often rely on an underlying model, whether explicitly constructed or implicitly assumed. However, what sets our framework apart is the introduction of an explicit \textit{interpretability function}, which provides a formal mechanism for extracting relevant metrics from the simplified model. This function dictates how the explanation should be constructed and evaluated, ensuring that the process is rigorous and systematic rather than ad hoc. 
As already mentioned, this article considers the weight costs obtained from of an inverse optimal control problem at the output of interpretability function. 
However, it is also discussed how alternative interpretability functions could be employed, establishing connections with other existing XAI methods and further generalizing the applicability of our framework.

The article describes in Section~\ref{sec:background} the main problem that we want to address in a formal way. Section~\ref{sec:interpretabilityfunctions} introduces different interpretability functions that can be applied to LTI systems. In particular, we emphasize the relevance of inverse optimal control methods in revealing the trade-offs that the black-box controller is effectively operating within its control policy. Section~\ref{sec:examples} provides several numerical illustrations demonstrating how our approach can highlight phenomena in controlled systems that are not immediately intuitive.

\section{Problem Statement}  
\label{sec:background}  

We consider a discrete-time, time-invariant dynamical system described by:  
\begin{subequations} \label{eq:plant}  
  \begin{align}  
   x(k+1) &= f(x(k), u(k)), \\  
   y(k) &= g(x(k), u(k)),     
  \end{align}  
\end{subequations}  
where $f$ and $g$ are unknown functions governing the system dynamics. Here, $x(k)$ represents the system state at time step $k$, while $u(k)$ and $y(k)$ denote the input and output signals, respectively. We assume that $u(k)$ is a controllable input, and we have access to experimental data in the form of input-output (I/O) pairs:  
\begin{equation}  
\mathcal{D}_{N_{data}} = \left\{ \left( u(k), y(k) \right) \right\}_{k=1}^{N_{data}},  
\end{equation}  
where $N_{data}$ is the number of collected data points.  

The goal of data-driven control is to directly synthesize a control law from the dataset $\mathcal{D}_{N_{data}}$, bypassing explicit identification of the system dynamics in \eqref{eq:plant}. Specifically, we seek a control law of the form:  
\begin{equation}  
    u(\cdot) = \phi(y(\cdot)),  
\end{equation}  
where $\phi$ is an operator mapping the measured output signal into the control signal to achieve a desired closed-loop behavior.  
Several approaches exist to achieve this without first identifying a model of the system. Prominent methodologies include:  

\begin{itemize}  
    \item \textit{Reinforcement Learning (RL)}: RL-based strategies learn an optimal control policy by interacting with the system and optimizing a performance criterion. Techniques such as \textit{Q-learning} \cite{sutton2018reinforcement} and \textit{policy gradient} methods \cite{recht2019tour} have demonstrated effectiveness in high-dimensional and nonlinear settings, although they often require extensive exploration and suffer from sample inefficiency.  

    \item \textit{Willems' Fundamental Lemma}: Originally formulated for linear time-invariant (LTI) systems \cite{willems2005note}, this method enables direct control design using raw data. Recent works have extended this concept to certain nonlinear systems \cite{molodchyk2024exploring}.  

    \item \textit{Model Reference Approaches}: These methods, often used in adaptive control \cite{narendra1989stable}, aim to shape the system’s behavior to match a desired reference model through data-driven techniques \cite{breschi2021direct,de2023data,busetto2025meta}. Some theoretical studies also compare direct data-driven approaches with traditional system identification, identifying conditions under which the former is preferable \cite{formentin2014comparison}.  

    \item \textit{Data-Driven Predictive Control}: Methods based on subspace identification \cite{hou2013model} use past data to predict future system behavior and compute control actions accordingly \cite{breschi2023data,chiuso2025harnessing}. Extensions incorporating Koopman operator theory \cite{korda2018linear} and Gaussian process regression \cite{hewing2020learning} further enhance predictive capabilities for nonlinear systems.  
\end{itemize}  

Although this list is not exhaustive, it highlights the increasing adoption of data-driven control techniques in real-world applications. As these methods become more prevalent, particularly in safety-critical domains such as autonomous vehicles, robotics, and healthcare, understanding the reasoning behind a controller’s decisions becomes crucial.  

\subsection{Explainability in Data-Driven Control}
\begin{figure}  
    \begin{center}  
        \begin{tikzpicture}[very thick]
   
    \node (method) at (3.9, 0.4) [draw, minimum width=5.5cm, minimum height=2.8cm, rounded corners=5, line width=0.75pt, fill=mypurple!20] {};
    \node (NN) at (0.2,0) [draw, minimum width=1.4cm, minimum height=1.5cm, fill=lightgray!70, rounded corners=3, line width=0.75pt] {};
    \node (model) at (2.3,0) [draw, minimum width=1.8cm, minimum height=1.5cm, fill=white, rounded corners=3, line width=0.75pt] {};
    \node (model_shifted) at (1.45,-0.2) [minimum width=1.8cm, minimum height=1.5cm, line width=0.75pt] {};
    \node (interpret) at (5.5,0) [draw, minimum width=1.8cm, minimum height=1.5cm, fill=white, rounded corners=3, line width=0.75pt] {};

    \node[draw, circle, minimum size=0.25cm, fill=white, line width=0.5pt, inner sep=0] (left_top)      at (-0.25,  0.5) {};
    \node[draw, circle, minimum size=0.25cm, fill=white, line width=0.5pt, inner sep=0] (left_middle)   at (-0.25,  0) {};
    \node[draw, circle, minimum size=0.25cm, fill=white, line width=0.5pt, inner sep=0] (left_bottom)   at (-0.25, -0.5) {};
    \node[draw, circle, minimum size=0.25cm, fill=white, line width=0.5pt, inner sep=0] (center_top)    at ( 0.2,  0.25) {};
    \node[draw, circle, minimum size=0.25cm, fill=white, line width=0.5pt, inner sep=0] (center_bottom) at ( 0.2, -0.25) {};
    \node[draw, circle, minimum size=0.25cm, fill=white, line width=0.5pt, inner sep=0] (right_top)     at ( 0.65,  0.5) {};
    \node[draw, circle, minimum size=0.25cm, fill=white, line width=0.5pt, inner sep=0] (right_middle)  at ( 0.65,  0) {};
    \node[draw, circle, minimum size=0.25cm, fill=white, line width=0.5pt, inner sep=0] (right_bottom)  at ( 0.65, -0.5) {};

    \draw [-stealth, line width=0.5] (center_top) -- (right_top);
    \draw [-stealth, line width=0.5] (center_top) -- (right_bottom);
    \draw [-stealth, line width=0.5] (center_bottom) -- (right_top);
    \draw [-stealth, line width=0.5] (center_bottom) -- (right_bottom);
    \draw [-stealth, line width=0.5] (center_top) -- (right_middle);
    \draw [-stealth, line width=0.5] (center_bottom) -- (right_middle);
    \draw [stealth-, line width=0.5] (center_top) -- (left_top);
    \draw [stealth-, line width=0.5] (center_top) -- (left_bottom);
    \draw [stealth-, line width=0.5] (center_bottom) -- (left_top);
    \draw [stealth-, line width=0.5] (center_bottom) -- (left_bottom);
    \draw [stealth-, line width=0.5] (center_top) -- (left_middle);
    \draw [stealth-, line width=0.5] (center_bottom) -- (left_middle);
    
    \begin{axis}[domain=-1.5:3, at={(model_shifted)}, width=2.5cm, height=2.3cm, axis lines=none, clip=false]
        \addplot [lightgray!50, mark size=8, only marks, forget plot] coordinates {(1, 0.841)};
        \addplot [myaqua, domain=0:2, samples=30, only marks, mark=*, mark size=0.25, forget plot] {sin(deg(\x)) + sin(360*rnd)*sqrt(-2*0.1*ln(rnd))}; 
        \addplot [samples=200, opacity=0, mark=none, line width=0.75, forget plot] {sin(deg(\x))};
    \end{axis}
    \begin{axis}[domain=-1.5:3, at={(model_shifted)}, width=2.5cm, height=2.35cm, axis lines=none, clip=false]
        \addplot [opacity=0, mark size=8, only marks, forget plot] coordinates {(1, 0.841)};
        \addplot [opacity=0, domain=0:2, samples=30, only marks, mark=*, mark size=0.35, forget plot] {sin(deg(\x)) + sin(360*rnd)*sqrt(-2*0.1*ln(rnd))};
        \addplot [samples=200, black, mark=none, line width=0.75, forget plot] {sin(deg(\x))};
        \addplot [myyellow, domain=-0.8:2.8, line width=1.25] {\x*0.45+0.45};
        \addplot [fill=myyellow, only marks, mark size=1.5, line width=0.5] coordinates {(1, 0.841)};
    \end{axis}

    \node (first_row)       at (2.625, -0.05) [draw, minimum width=0.15cm, minimum height=0.15cm, inner sep=0, fill=white, line width=0.5] {};
    \node (second_row_1)    at (2.375, -0.25) [draw, minimum width=0.15cm, minimum height=0.15cm, inner sep=0, fill=mygreen!65, line width=0.5] {\scalebox{.4}{\checkmark}};
    \node (second_row_2)    at (2.875, -0.25) [draw, minimum width=0.15cm, minimum height=0.15cm, inner sep=0, fill=myred!65, line width=0.5] {\scalebox{.4}{$\times$}};
    \node (third_row_1)     at (2.25, -0.55) [draw, minimum width=0.15cm, minimum height=0.15cm, inner sep=0, fill=mygreen!65, line width=0.5] {\scalebox{.4}{\checkmark}};
    \node (third_row_2)     at (2.5, -0.55) [draw, minimum width=0.15cm, minimum height=0.15cm, inner sep=0, fill=myred!65, line width=0.5] {\scalebox{.4}{$\times$}};
    \node (third_row_3)     at (2.75, -0.55) [draw, minimum width=0.15cm, minimum height=0.15cm, inner sep=0, fill=mygreen!65, line width=0.5] {\scalebox{.4}{\checkmark}};
    \node (third_row_4)     at (3, -0.55) [draw, minimum width=0.15cm, minimum height=0.15cm, inner sep=0, fill=myred!65, line width=0.5]   {\scalebox{.4}{$\times$}};
    \draw [line width=0.5] (first_row) -- (2.625, -0.25) -- (second_row_1);
    \draw [line width=0.5] (2.625, -0.25) -- (second_row_2);
    \draw [line width=0.5] (second_row_1) -- (2.375, -0.4) -- (2.25, -0.4) -- (third_row_1);
    \draw [line width=0.5] (2.375, -0.4) -- (2.5, -0.4) -- (third_row_2);
    \draw [line width=0.5] (second_row_2) -- (2.875, -0.4) -- (2.75, -0.4) -- (third_row_3);
    \draw [line width=0.5] (2.875, -0.4) -- (3, -0.4) -- (third_row_4);
    
    \node [draw, lightgray, circle, minimum size=0.7cm, line width=1pt, inner sep=0] (right_bottom)  at ( 5.5, 0) {};
    \draw [-stealth, line width=0.5] (5.5, -0.55) -- (5.5, 0.6);
    \draw [-stealth, line width=0.5] (4.85, 0) -- (6.2, 0);
    \node [draw, myblue, circle, minimum size=0.1cm, line width=1pt, inner sep=0] (right_bottom)  at ( 5.6, 0.2) {};
    \node [draw, myblue, circle, minimum size=0.1cm, line width=1pt, inner sep=0] (right_bottom)  at ( 5.6, -0.2) {};
    \draw [myred, line width=1] (5.3, -0.05) -- (5.4, 0.05);
    \draw [myred, line width=1] (5.4, -0.05) -- (5.3, 0.05);
    \draw [myred, line width=1] (5.75, -0.05) -- (5.85, 0.05);
    \draw [myred, line width=1] (5.85, -0.05) -- (5.75, 0.05);
    \node at (6.1, -0.2) {\scriptsize $\mathfrak{Re}$};
    \node at (5.2, 0.5) {\scriptsize $\mathfrak{Im}$};

    \draw [->, line width=0.75] (NN) -- (model);
    \draw [->, line width=0.75] (interpret) -- (7.5, 0);
    \draw [->, line width=0.75] (model) -- (interpret);
        
    \node at (0.2, 1)       {\footnotesize Black-Box};
    \node at (2.3, 1.25)    {\footnotesize Underlying};
    \node at (2.3, 0.95)    {\footnotesize Modeling};
    \node at (3.9, 0.6)     {\footnotesize Model};
    \node at (3.9, 0.3)     {\footnotesize Parameters};
    \node at (5.5, 1.25)    {\footnotesize Interpretability};
    \node at (5.5, 0.95)    {\footnotesize Function};
    \node at (7.4, 0.3)     {\footnotesize Explanation};
    \node at (3.9, 1.8)     [draw, fill=mypurple!10, rounded corners=3, line width=0.75pt] {Explanation Method};
    
\end{tikzpicture}
    \end{center}  
    \caption{\it Schematic formal representation of an explanation method.}  
    \label{fig:SysIDExplanation}  
\end{figure}
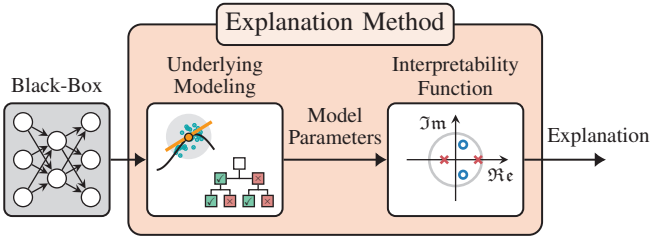

In this work, we tackle the explainability problem for data-driven controllers following a two-step approach, as illustrated in Figure~\ref{fig:SysIDExplanation}.
The first step involves selecting an underlying explanatory model from an interpretable class of models.
This model is not required to capture the full complexity of the system, but should instead provide a simplified yet insightful representation. Indeed, the choice of the model class should be guided by interpretability as the primary criterion.
Once this modeling procedure is established, the second step involves deriving an explanation by mapping the parameters of the simplified model into quantitative measures that assess the internal mechanisms of the original complex system. This mapping is formalized through what we refer to as the ``interpretability function''.

Although not always explicitly acknowledged in the XAI literature, many widely used XAI methods inherently follow this two-step approach. For example, LIME typically employs linear regression models as its interpretable class and provides the regression coefficients as explanations, effectively using the identity function as its interpretability function \cite{biparva2023interpretation,biparva2023application}.
Similarly, as shown in \cite{biparva2024incorporating}, SHAP constructs explanations using linear structural equation models as the interpretable class, computing them across all $n!$ possible permutations of the $n$ input variables. The explanation is then derived by averaging input contributions over these permutations, thereby employing an interpretability function that is effectively an average over all possible structural equation models.

The observation that several XAI methods follow this structure can contribute to a better formalization of what it means to explain a black-box model and has the potential to inspire the development of novel XAI tools that can be systematically compared within the same framework. However, these considerations fall outside the scope of this work. Instead, our focus is on proposing an explanatory tool specifically for feedback controllers.
Such a tool needs to convey a more \textit{nuanced} understanding of the closed-loop structure. Indeed, while standard XAI methods typically explain static relationships, they often fail to account for the time-interdependence between the controller's decisions and the plant dynamics. As a result, they overlook the dynamic and feedback-driven nature of control systems, which fundamentally distinguishes control problems from traditional machine learning tasks.

In the following section, we motivate our specific choices for the underlying class of models and the interpretability function, highlighting how these selections align with the unique challenges of explaining data-driven controllers in dynamic systems.

\section{Inverse optimal control as an interpretability function}
\label{sec:interpretabilityfunctions}
In this section, we propose a specific tool for the explainability of feedback controllers. Following the approach delineated in the previous section, we need to select a class of interpretable models and an interpretability function that can translate the parameters of these models into meaningful quantities capable of representing the behavior of the controller within the feedback loop.
In this respect, we choose 
\begin{itemize}
    \item   Linear Time-Invariant (LTI) systems as our class of interpretable models
    \item   the solution of an inverse Linear Quadratic Regulator (LQR) based on the linearized dynamics of the controller as our interpretable function.   
\end{itemize}
and motivate these choices in the following.

\subsection{LTI systems as class of interpretable models}
While nonlinear systems exhibit complex dynamics, meaningful insights can still be gained by analyzing interpretable approximations of their local behavior.
A natural choice for the class of interpretable models is given by Linear Time-Invariant (LTI) systems. They are mathematically tractable, with a rich theoretical foundation that enables precise analysis of system properties such as stability, controllability, and observability. 
Furthemore, using LTI models ensures consistency with existing explainable AI methods, many of which rely on linear approximations (e.g., LIME), making the application of these methods to dynamic systems more coherent and effective.


\subsection{Interpretability functions for LTI systems}
The choice of linearization of the controller's dynamics as the underlying modeling procedure enables the selection of several natural interpretability functions, which include:
\begin{itemize}
    \item \emph{Difference Equation Coefficients.} 
    Given the linearized behavior of the controller, the interpretability function can be chosen to provide the coefficients of the numerator and denominator of the corresponding transfer function, offering insights into the system's input-output relationships and its response to control inputs. These coefficients also provide a natural way to describe the intertemporal relationships among the different time signals of the system.

    \item \emph{Pole-Zero Representation.} An alternative (and equivalent) interpretability function could be given by the system’s poles and zeros. Poles provide critical information about stability and transient response, while zeros influence controllability and system performance. Examining how these characteristics change under different conditions can offer valuable insights into the controller’s impact on system behavior especially in terms of time constants and time separations.

    \item \emph{Inverse Optimal Control.} A more nuanced yet powerful interpretability function involves the use of inverse optimal control techniques, to infer a cost function that explains the controller’s local behavior. 
    This approach enables a \textit{structured interpretation of control strategies} in terms of trade-offs between output error and control effort. Moreover, it provides a systematic way to analyze controllers that were tuned empirically or derived from learning-based methods, revealing hidden objectives that may not have been explicitly defined.
\end{itemize}

The first two interpretability functions are equivalent, as one can be obtained directly from the other.
They offer the advantage of being an immediate dynamic extension of existing XAI techniques, providing a straightforward means to analyze the controller’s input-output relationships and stability characteristics.
On the other hand, the approach based on inverse optimal control is definitely control-specific and connects deeply into performance-related aspects of the system.
It can provide valuable information on the interrelations between signals in a way that directly reflects the controller’s objectives and priorities.
Furthermore, the inverse optimal control approach has the potential to generalize to other model classes beyond linear systems, making it a versatile tool for explaining feedback controllers in a broader range of applications. 
For these reasons, this article explores a method based on inverse optimal control as its interpretability function.

\subsection{Inverse LQR for explainability}
\label{sec:ilq}
Since the interpretable class of models is given by LTI systems, the most natural approach to solving an inverse optimal control problem is by considering the standard formulation of inverse Linear Quadratic Regulator (LQR) optimal control.
The conditions for the existence of a solution to such a problem in the case of linear dynamics and a stabilizing controller are quite general \cite{priess2014solutions} guaranteeing the existence of the explanation. 
The mathematical details of the numerical methods employed in this paper to compute the inverse LQR problem are inspired from \cite{zhang2019inverse} and described in the following.

The standard finite-horizon LQR problem is given by
\begin{align}\label{eq:lqr_problem}
    \min_{x, u} & \; x(N)^\top S x(N) + \sum_{k=1}^{N-1} (u(k)^\top R u(k) + x(k)^\top Q x(k)) \\
    \textrm{s.t.} & \; x(k+1)= A x(k) + B u(k), \quad x(1) = \bar{x}, \label{eq_state_equation}
\end{align}
where $S, Q \succeq 0$, and $R \succ 0$.
Such inverse optimal control problem seeks to determine $(S, Q, R)$ given $(A, B)$ and observed optimal trajectories or control inputs, potentially with noise\footnote{If the system matrices are not available, they can be estimated from the same data using identification techniques, such as subspace methods.}. In this work, we assume $S = 0$ for simplicity. Nonetheless, the problem remains well-posed as $Q$ is uniquely determined for a given closed-loop system \cite{zhang2019inverse}.

Assume to have access to $M$ sets of state-input trajectories $\{\mathcal{D}_N^{(i)}\}_{i=1}^{M}$ generated by the data-driven controller\footnote{These datasets can be obtained by simulating the system in closed-loop.}, namely $\mathcal{D}_N^{(i)} = \{(x_d^{(i)}(k), u_d^{(i)}(k))\}_{k=1}^{N}$, where the subscript $d$ indicates that data is affected by noise. To explain each collected closed-loop trajectory, we leverage the Pontryagin Maximum Principle (PMP), which provides necessary and sufficient conditions for optimality through the existence of an adjoint variable $\lambda$ that satisfies the equations:
\begin{subequations}\label{eq:PMP}
    \begin{align}
        \lambda(k) &= A^\top \lambda(k+1) + Q x(k), \quad k = 2, \dots, N-1, \label{eq:lambda_update} \\
        \lambda(N) &= 0, \\
        u(k) &= -R^{-1} B^\top \lambda(k+1), \quad k = 1, \dots, N-1. \label{eq:input_equation}
    \end{align}
\end{subequations}
Using these conditions, we can search for the matrices $Q$ and $R$ that satisfy the PMP given the collected datasets $\mathcal{D}_N^{(i)}$. 
Since each dataset is noisy, we cannot use the measured state $x_d^{(i)}$ to solve \eqref{eq:PMP}. Instead, we estimate the noiseless state sequence by minimizing the difference between the predicted evolution $x^{(i)}$ and the measured trajectories $x_d^{(i)}$. To this end, considering the state equation $\eqref{eq_state_equation}$ and the PMP input equation \eqref{eq:input_equation}, we write the noiseless state dynamics as
\begin{align}\label{eq:PMP_state_lambda}
    x(k+1) &= A x(k) - B R^{-1} B^\top \lambda(k+1).
\end{align}
Through \eqref{eq:PMP_state_lambda}, the state dynamics can be reconstructed from an initial condition $\bar{x}$, the cost function matrices $Q$, $R$, and the adjoint variable update \eqref{eq:lambda_update}. Therefore, we can obtain estimates $\hat{Q}$, $\hat{R}$ of the cost function matrices by taking the minimum state reconstruction error over each collected data, namely solving the optimization problem
\begin{subequations}\label{eq:IOC_problem}
    \begin{align}
        \min_{Q,R} & \quad \frac{1}{M} \sum_{i=1}^{M} \sum_{k=2}^{N} \left\lVert x_d^{(i)}(k) - x^{(i)}(k) \right\rVert^2 \\
        \textrm{s.t.} & \quad \lambda^{(i)}(k) = A^\top \lambda^{(i)}(k\!+\!1) + Q x^{(i)}(k), \\
        & \quad x^{(i)}(k\!+\!1) = A x^{(i)}(k) - B R^{-1} B^\top \lambda^{(i)}(k\!+\!1), \\
        & \quad \lambda^{(i)}(N) = 0, \\
        & \quad x^{(i)}(1) = \bar{x}^{(i)},
    \end{align}
\end{subequations}
which is convex and can be solved using standard nonlinear optimization solvers. Moreover, for $M \to \infty$, the estimates $\hat{Q}$, $\hat{R}$ of the IOC problem \eqref{eq:IOC_problem} converge to the true values of the forward problem \eqref{eq:lqr_problem} \cite{zhang2019inverse}.

\section{Numerical examples}
\label{sec:examples}

In this section, we apply the IOC explainability framework to two case studies: a second-order LTI system and an inverted pendulum. 
To generate the trajectories required for solving the IOC optimization problem \eqref{eq:IOC_problem}, we simulate the systems in closed-loop with a model predictive controller (MPC). While an MPC is not a data-driven controller, we assume the observed closed-loop I/O trajectories originate from an unknown control strategy that we aim to explain.
From this perspective, this approach allows us to compare the generated explanations $\hat{Q}$, $\hat{R}$ with a ground truth provided by the control hyperparameters $Q_{MPC}$, $R_{MPC}$ used to tune the MPC.

\subsection{Second-order LTI system}
\begin{figure}[t]
    \centering
    \begin{subfigure}[t]{\linewidth}
        \centering
        \includegraphics[width=0.8\linewidth]{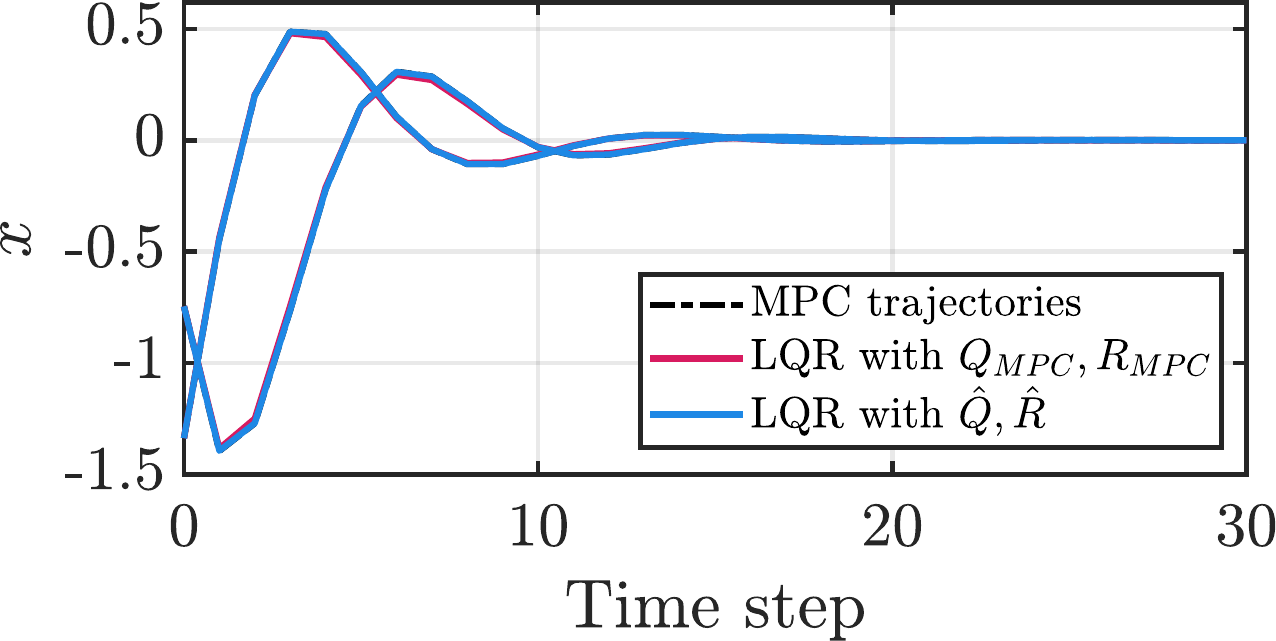}
        \caption{$T=10$.}
        \label{fig:long_prediction_horizon}
    \end{subfigure}
    \\[2mm]
    \begin{subfigure}[t]{\linewidth}
        \centering
        \includegraphics[width=0.8\linewidth]{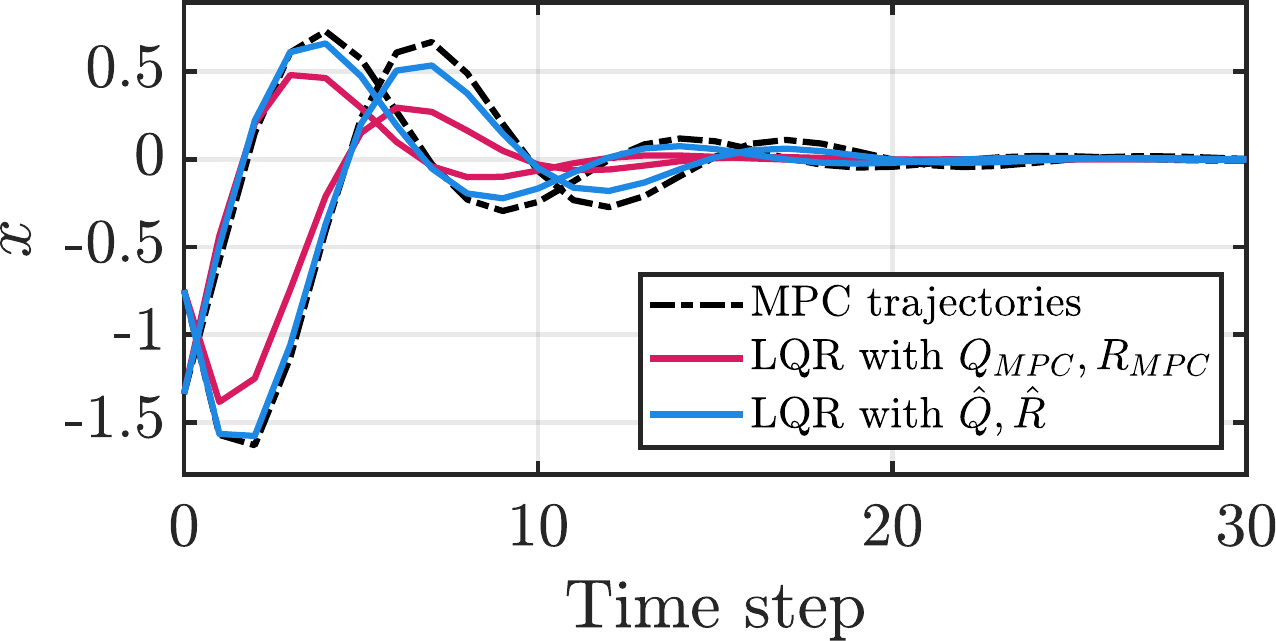}
        \caption{$T=5$.}
        \label{fig:short_prediction_horizon}
    \end{subfigure}
    \caption{\it LQR trajectories using MPC weights $Q_{MPC}$, $R_{MPC}$ and using IOC weights $\hat{Q}$, $\hat{R}$.}
    \label{fig:prediction_horizon_comparison}
\end{figure}

Consider the fully measurable second-order system 
\begin{align*}  
   x(k+1) & = Ax(k) + Bu(k), \\  
   y(k) & = x(k) + e(k),   
\end{align*}  
where $e$ is a zero-mean white measurement noise with covariance $\sigma^2I$, $\sigma = 0.01$, and
\begin{equation*}
    A = \begin{bmatrix}
        1 & 1 \\ -0.5 & 1
    \end{bmatrix}, \quad
    B = \begin{bmatrix}
        0.5 & 0 \\ 0 & 0.5
    \end{bmatrix}.
\end{equation*}
The system is controlled using an unconstrained quadratic MPC that regulates the state to zero with the cost function weights matrices
\begin{equation*}
    Q_{MPC} = \begin{bmatrix}
        0.3 & 0 \\ 0 & 0.1
    \end{bmatrix}, \quad
    R_{MPC} = \begin{bmatrix}
        1 & 0 \\ 0 & 1
    \end{bmatrix}.
\end{equation*}
Our objective is to explain the behavior of the MPC for different prediction horizons $T$. To this end, we conduct two experiments with $T=5$ and $T=10$. In each experiment, we collect $M=30$ closed-loop trajectories of length $N=30$ by sampling the initial conditions $x(0)$ from a multivariate uniform distribution over $[-2, 2] \times [-2, 2]$. Then, we compute an equivalent LQR-based explanation $(\hat{Q}, \hat{R})$ through the IOC algorithm in \eqref{eq:IOC_problem}, yielding the following matrices:
\begin{align*}
    \hat{Q}_{T=10} & = \begin{bmatrix}
        0.265 & 0.015 \\ 0.015 & 0.112
    \end{bmatrix}, &  
    \hat{R}_{T=10} & = \begin{bmatrix}
        0.988 & 0 \\ 0 & 1.012
    \end{bmatrix}, \\
    \hat{Q}_{T=5} & = \begin{bmatrix}
        0.014 & -0.02 \\ -0.02 & 0.041
    \end{bmatrix}, &  
    \hat{R}_{T=5} & = \begin{bmatrix}
        1.06 & 0 \\ 0 & 0.949
    \end{bmatrix}.
\end{align*}
For $T=10$, the obtained explanation resembles the original MPC weights; however, for $T=5$, the explanation matrices show a different trend. In particular, the magnitude of $\hat{Q}_{T=5}$ compared to $\hat{R}_{T=5}$ is smaller, indicating a less aggressive control strategy compared to $T=10$ (see Figure \ref{fig:prediction_horizon_comparison} for a comparison between $T=5$ and $T=10$). 

Moreover, while $Q_{MPC}$ assigns a larger penalty to the first state, the explanation $\hat{Q}_{T=5}$ weights more the second state. This result indicates that, \textit{over the full simulation length} $N=30$, the LQR that best approximates the measured closed-loop behavior is characterized by weight matrices which differ in nature from than those used in the original MPC formulation.

To validate this claim, Figure~\ref{fig:prediction_horizon_comparison} shows the MPC trajectories compared to the LQR trajectories computed both using $Q_{MPC}, R_{MPC}$ and $\hat{Q}, \hat{R}$. When $T=5$, the trajectories obtained with $\hat{Q}, \hat{R}$ match more closely the MPC trajectories.

\subsection{Inverted pendulum}
Next, we consider the fully measurable inverted pendulum \cite{Gross:22a} with dynamics
\begin{align*}
    \theta(k+1) & = \theta(k) + \tau \omega(k), \\
    \omega(k+1) & = \frac{\tau g}{l} \sin(\theta(k)) + \left( 1- \frac{
    \tau d}{ml^2}\right) \omega(k) + \frac{\tau}{ml^2} u(k),
\end{align*}
where $\theta$ and $\omega$ are the angular position and speed, respectively, and $u$ is the control input. The system's parameters are listed in Table~\ref{tab:pendulum_parameters}. 

\begin{table}[t]
    \centering
    \begin{tabular}{c|c|c|c|c|c}
        \textbf{Parameter [unit]} & $m$ [kg] & $g$ [m/s\textsuperscript{2}] & $l$ [m] & $d$ [Nms] & $\tau$ [s] \\
        \textbf{Value} & 0.676 & 9.81 & 0.45 & 0.1 & 0.02
    \end{tabular}
    \caption{Parameters of the inverted pendulum.}
    \label{tab:pendulum_parameters}
\end{table}

Since the pendulum dynamics are nonlinear and the IOC algorithm assumes an underlying linear system, we focus on explaining the local behavior of the pendulum around a given angular position $\bar{\theta}$. To do so, we collect $M=30$ closed-loop trajectories of length $N=50$ in the neighborhood of $\bar{\theta}$ using an unconstrained nonlinear MPC (NMPC) with prediction horizon $T=5$. The NMPC cost function is quadratic and penalizes the deviation of $\theta(k)$ and $u(k)$ from the reference values $\theta_{r}(k)$, $u_r(k)$, with $\theta_r(k) = \bar{\theta} + e(k)$, where $e \sim\mathcal{N}(0,0.05^2)$ is a Gaussian perturbation that excites the system, and $u_r(k)$ is the equilibrium input corresponding to $\theta_r(k)$. The NMPC cost function matrices are 
\begin{equation*}
    Q_{MPC} = \begin{bmatrix}
        1000 & 0 \\ 0 & 100
    \end{bmatrix}, \quad
    R_{MPC} = 1,
\end{equation*}
and the initial conditions $\theta(0)$, $\omega(0)$ are sampled from a multivariate uniform distribution over $[\bar{\theta}-0.3, \bar{\theta}+0.3] \times [-0.3, 0.3]$.

With the measured trajectories, we compute a least-squares estimate of the linearized system matrices $A(\bar{\theta}), B(\bar{\theta})$ and then solve the IOC problem. Since the pendulum dynamics are nonlinear, the behavior of the NMPC, and consequently its explanation, depends on the operating point of the system. For this reason, we study the explanations for the two different angular positions $\bar{\theta}_1 = 0$~rad (upright position) and $\bar{\theta}_2 = \pi/2$~rad (horizontal position).

The explanations obtained by the IOC algorithm for the two angles $\bar{\theta}_1$, $\bar{\theta}_2$ are
\begin{equation*}
    \hat{Q}_{\bar{\theta}_1} = \begin{bmatrix}
        11.674 & 18.179 \\ 18.179 & 28.309
    \end{bmatrix}, \quad 
    \hat{Q}_{\bar{\theta}_2} = \begin{bmatrix}
        12.949 & 12.408 \\ 12.408 & 19.183
    \end{bmatrix},
\end{equation*}
where $\hat{R}_{\bar{\theta}_1} = \hat{R}_{\bar{\theta}_2} =  R_{MPC} = 1$ is forced by the optimization problem\footnote{If $\hat{Q}$, $\hat{R}$ are solutions of \eqref{eq:IOC_problem}, then, for any $\alpha>0$, also $\alpha\hat{Q}$, $\alpha\hat{R}$ are solutions of \eqref{eq:IOC_problem}. Therefore, we impose $\hat{R}=1$ to obtain a unique solution of the IOC.}. 
Comparing $\hat{Q}_{\bar{\theta}_1}$ and $\hat{Q}_{\bar{\theta}_2}$, we observe that the weights are larger for the operating point $\bar{\theta}_1=0$~rad. Since the IOC computes the optimal ratio between the weight matrices $Q$ and $R$, larger weights in $\hat{Q}_{\bar{\theta}_1}$ imply that the NMPC penalizes the control input less when the pendulum is upright. This behavior is coherent with a physical understanding of the pendulum dynamics, as a larger input is required to keep the pendulum horizontal.
We observe that, in this case, the matrices providing the explanation exhibit non-negligible off-diagonal components, indicating that the controller is optimizing a cost function where the state components are significantly coupled. This contrasts with the nominal cost function used in the MPC, which is diagonal, suggesting that the learned controller accounts for interactions between different state variables rather than treating them independently.

\begin{figure}[ht]
    \centering
    \includegraphics[width=0.99\linewidth]{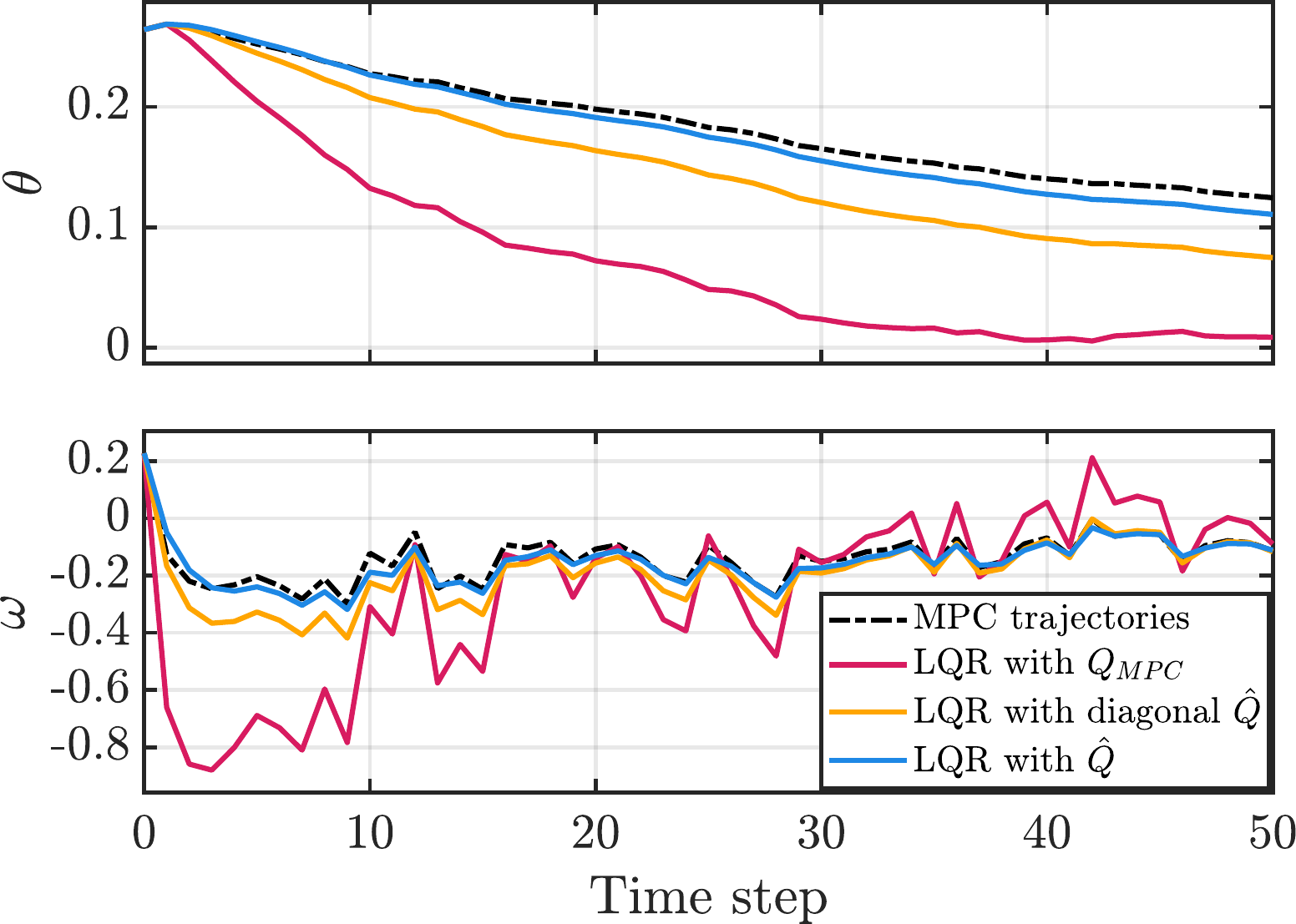}
    \caption{\it MPC and LQR trajectories for the pendulum near the equilibrium point $\bar{\theta}_1 = 0$ rad.}
    \label{fig:pendulum}
\end{figure}

In order to validate this point, we can attempt to solve the optimization problem \eqref{eq:IOC_problem} while enforcing the matrices to be diagonal, leading to
\begin{equation*}
    \hat{Q}_{\bar{\theta}_1, d} = \begin{bmatrix}
        63.901 & 0 \\ 0 & 49.001
    \end{bmatrix}, \quad 
    \hat{Q}_{\bar{\theta}_2, d} = \begin{bmatrix}
        29.125 & 0 \\ 0 & 19.578
    \end{bmatrix}.
\end{equation*}
Figure \ref{fig:pendulum} shows a comparison between the LQR trajectories computed using the MPC weight $Q_{MPC}$, the non-diagonal explanations $\hat{Q}$ and the diagonal explanation $\hat{Q}_d$ for $\bar{\theta}_1$, highlighting that the adoption of a non-diagonal $\hat Q$ maximizes adherence to the measured trajectories, especially in transient, and hence provides a more accurate description of the actual cost being optimized by the controller.

\section{Concluding remarks}
\label{sec:conclusions}

Developing explainability and interpretability tools for controllers designed through AI-based or data-driven methods is essential to enhance transparency, enable debugging, validate behavior, and ensure compliance with emerging legal regulations.
However, explaining the behavior of dynamic controllers presents unique challenges, particularly due to their temporal dependencies and interactions within closed-loop systems.
To tackle these challenges, this article introduces a novel methodology that leverages standard control-theoretic tools to extract information on the behavior of such controllers.
We propose a general framework where explanations are derived via a reduced-complexity modeling process, followed by an interpretability function. Specifically, we demonstrate a tool where the reduced-complexity model is obtained by linearizing the system around the trajectory to be explained, and the interpretability function is constructed from the cost weights of a linear quadratic problem. Through several examples, we show that the explanations provided by this approach offer insights into phenomena that might otherwise be seen as counterintuitive or reveal unintended effects that could be commonly encountered in the design of data-driven controllers. 

Future work could focus on extending the methodology to nonlinear systems, where linear approximations may be insufficient. Enhancing scalability for real-time applications and high-dimensional state spaces is another key direction. Additionally, integrating domain-specific knowledge could further improve the interpretability of controllers in critical areas such as autonomous driving, robotics, and energy management.


\bibliographystyle{ieeetran}
\bibliography{main}


\end{document}